\documentclass[%
 preprint,
superscriptaddress,
 amsmath,amssymb,
 aps, physrev,
 showkeys,
hidelinks
]{revtex4-2}

\usepackage{graphicx}
\usepackage{dcolumn}
\usepackage{bm}
\usepackage{hyperref}
\usepackage{physics}
\usepackage{xcolor}
\usepackage{xr-hyper}
\newcommand{\im}{\mathrm{i}}
\newcommand{\eV}{\,\mathrm{eV}}
\newcommand{\fs}{\,\mathrm{fs}}
\newcommand{\wc}{\omega_\mathrm{c}}

\newcommand{\ee}{\mathrm{e}}
\newcommand{\g}{\mathrm{g}}

\begin{document}

\title{\textbf{Second-order coherence properties of ultrafast polariton dynamics in plasmonic lattices} 
}%

\author{Evgeny A.~Mamonov}
\author{Lukas Freter}
\author{Sioneh Eyvazi}
\affiliation{%
Department of Applied Physics, Aalto University School of Science, Aalto FI-00076, Finland
}%
\author{Elliot W.~Lloyd}
\affiliation{SUPA, School of Physics and Astronomy, University of St Andrews, St Andrews, KY16 9SS, UK}
\author{Päivi Törmä}%
 \email{Contact author: paivi.torma@aalto.fi}
\affiliation{%
Department of Applied Physics, Aalto University School of Science, Aalto FI-00076, Finland
}%

\date{\today}

\begin{abstract}

The first-order coherence properties of strongly coupled systems supporting polariton lasing and Bose-Einstein condensation (BEC) have been thoroughly studied and show consistent results. In contrast, second-order coherence properties have been found to vary for different systems, and the second-order coherence function can deviate from the value of one, typical for atomic BEC, revealing super-Poissonian fluctuations even above the BEC/polariton lasing threshold. This calls for a deeper understanding of the emission statistics and dynamics in light-based condensates, especially in the non-equilibrium regime. Here we demonstrate a coherent state of polariton emission from a plasmonic lattice measured as the second-order coherence function value separately for $\mathrm{\Gamma}$- and non-$\mathrm{\Gamma}$-point radiation (ground state and high-energy tail). We observe a high degree of coherence with the value $g^{(2)}(\tau=0)-1<10^{-4}$. We also demonstrate the ultrafast ($>$1 THz) nature of the dynamics. The process can be interpreted as a BEC with ultrafast, sub-50-fs thermalization, but we also suggest an alternative explanation as superradiance in a system where the emitters have vibrational degrees of freedom.
\end{abstract}

\maketitle

\section{\label{sec:intro}Introduction}
The first-order coherence properties of different types of BECs and polariton condensates have been widely studied. Large-scale spatial and temporal coherence has been demonstrated for atomic systems \cite{pitaevskii_boseeinstein_2016, ketterle_coh, Bloch2000, PhysRevA.84.021605}, and different types of decay have been observed for photonic and polaritonic systems \cite{PhysRevB.90.205430, Baboux:18, Whittaker_2009, Damm2017, Marelic_2016}. Second-order coherence properties of various condensates have been studied less extensively. Usually, the main focus is placed on the temporal second-order coherence function and its value at zero time delay, $g^{(2)}(\tau=0)$. This value provides direct information about the statistics of the BEC, which is related to fluctuations in the number of condensed particles. Atomic BEC systems are described as canonical ensembles; therefore, they have $g^{(2)}(\tau=0)=1$, indicating a lack of particle-number fluctuations \cite{pitaevskii_boseeinstein_2016, schellekensHanburyBrownTwiss2005, PhysRevA.56.3291_br_rep, ottlCorrelationsCountingStatistics2005, hodgmanDirectMeasurementLongRange2011}. The same result is expected for lasing \cite{scullyQuantumOptics1997, ulrichPhotonStatisticsSemiconductor2007}. For photonic and polaritonic condensates, the situation is more complicated. In both cases, light emission in the BEC system starts from the fluorescence of the active medium below the BEC threshold, which has thermal statistics and $g^{(2)}(\tau=0)=2$. Then, at and above the threshold, some suppression of fluctuations is expected. In the case of photon BEC (a dye-filled microcavity in the weak coupling regime), one can effectively tune the molecular reservoir size by dye concentration and detuning the cavity frequency from the zero-phonon line of the dye, so that the second-order coherence value approaches one at different rates as the condensate fraction increases. Moreover, the emergence of second-order coherence need not coincide with the BEC threshold, making the grand-canonical ensemble an adequate description of the system in certain cases \cite{PhysRevLett.112.030401.grand-canonical, PhysRevLett.108.160403.grand-canonical_theory}.

Polariton BECs in the strong light-matter coupling regime are intrinsically at nonequilibrium and driven–dissipative \cite{Bloch2022_review, keeling_boseeinstein_2020}. Polariton–polariton interactions can introduce additional fluctuations into the emission properties. In these systems, a transition of $g^{(2)}(\tau=0)$ from two to one is also expected for the ground mode \cite{doanCoherenceCondensedMicrocavity2008, kavokin_microcavities_2017}, but accounting for polariton scattering raises the minimum achievable deviation from one \cite{PhysRevB.77.085317_theor_stat, sarchiEffectsNoiseDifferent2008}. Similar results have also been shown experimentally \cite{assmannPolaritonCondensatesHighly2011} and theoretically \cite{PhysRevB.78.073404}. For inorganic condensate systems, a comprehensive study of $g^{(2)}(\tau=0)$ for real-space fluctuations was reported in Ref.~\cite{Estrecho2018}: BEC systems with different detuning values were studied, and light-weight photonic polaritons for negative detuning and heavy excitonic polaritons for positive detuning were observed. Due to more efficient phonon-assisted relaxation for excitonic polaritons, which leads to fluctuation suppression, the second-order coherence function approached one for positive detuning, whereas it remained close to two even at high pump powers ($> 5\,P_{\mathrm{thr}}$) for negative detuning. A study of real-space second-order coherence in an organic BEC system with negative detuning \cite{PhysRevB.Anton} showed the absence of fluctuations in the overall number of polaritons in the condensate despite both static and dynamic disorder. The effect of dynamic disorder was demonstrated in Ref.~\cite{doi:10.1021/acsphotonics.7b00283} experimentally and theoretically by considering scattering between condensed and reservoir polaritons. A theoretical study of such a system~\cite{shishkovExactAnalyticalSolution2022} in the limit of fast thermalization and weak polariton interactions showed a transition of $g^{(2)}(\tau=0)$ to one above the BEC threshold for the ground mode and an anticorrelation between ground-mode emission and the thermalized high-energy distribution near threshold.

In principle, polariton emission in (organic) strongly coupled systems may also exhibit superradiance. Superradiance, as first introduced by Dicke in 1954~\cite{dickeCoherenceSpontaneousRadiation1954}, describes the collective spontaneous emission of an ensemble of $N$ closely spaced emitters in free space. Defining features of superradiance are an emitted peak intensity that scales as $N^2$ and a pulse width that vanishes as $N^{-1}$~\cite{grossSuperradianceEssayTheory1982}. Bonifacio and Lugiato~\cite{bonifacioCooperativeRadiationProcesses1975} introduced the term superfluorescence, which results from an initially incoherent system, typically a fully inverted ensemble of emitters. In contrast, the term superradiance is used for collective emission with some initial coherence in the system.
Superfluorescence was first observed in cold atom gases~\cite{skribanowitzObservationDickeSuperradiance1973, gibbsSinglePulseSuperfluorescenceCesium1977} and subsequently in solid state KCl crystals doped with oxygen~\cite{florianSuperradianceHighgainMirrorless1982,florianTimeresolvingExperimentsDicke1984}. Recently, superfluorescence was also observed in inorganic semiconductors such as CuCl quantum dots~\cite{miyajimaSuperfluorescentPulsedEmission2009}, bulk ZnTe~\cite{daiObservationSuperfluorescenceQuantum2011}, InGaAs quantum wells~\cite{timothynoeiiGiantSuperfluorescentBursts2012}, in thin film perovskite quantum dots~\cite{pooniaSuperfluorescenceElectronHolePlasma2024, aggarwalRoomTemperatureSuperfluorescence2025}, and perovskite superlattices~\cite{rainoSuperfluorescenceLeadHalide2018}. It was also observed in organic-inorganic hybrid perovskites~\cite{findikHightemperatureSuperfluorescenceMethyl2021, bilirogluRoomtemperatureSuperfluorescenceHybrid2022}. Moreover, cavity-assisted superfluorescence in perovskites was reported~\cite{zhouCooperativeExcitonicQuantum2020}, and a transition from superfluorescence to polariton condensation in a perovskite thin film on top of a Bragg reflector~\cite{maoObservationTransitionSuperfluorescence2024}. 

 In contrast to superfluorescence, a large effort has been put into studying ``excitonic superradiance", which describes the enhanced decay rate of excitons due to the coherent nature of the excitons in a two-dimensional structure~\cite{hanamuraRapidRadiativeDecay1988,itohQuantumConfinementExcitons1990, dingSuperradianceExcitonsSingle2008} or the radiative coupling of excitons in different quantum wells leading to enhanced reflectivity and line broadening~\cite{feldmannLinewidthDependenceRadiative1987,hubnerCollectiveEffectsExcitons1996a,hubnerOpticalLatticesAchieved1999,ammerlahnCollectiveRadiativeDecay2000,pozinaSuperradiantModeInAs2015,ivanovResonantOpticalReflection2023,shimosakoMultipleQuantumwellPolaritons2025}.
A similar enhancement of radiative decay has been observed for Frenkel excitons in organic J-Aggregates~\cite{deboerDephasinginducedDampingSuperradiant1990, fidderSuperradiantEmissionOptical1990}, R-phycoerythrin~\cite{wangSuperradianceHighDensity1995}, H-Aggregates~\cite{meinardiSuperradianceMolecularAggregates2003}, light-harvesting nanotubes~\cite{doriaPhotochemicalControlExciton2018}, and in J-Aggregates coupled to a metasurface~\cite{marangiEnhancingCooperativityMolecular2024}.
For a review of superradiance and superfluorescence in solid state systems, see Ref.~\cite{congDickeSuperradianceSolids2016}.

The second-order coherence of superradiance and superfluorescence has been studied both theoretically and experimentally. Early theoretical studies showed that $g^{(2)}(\tau=0)=2$ at $t=0$ for superfluorescence, but it decreases during the emission process \cite{bonifacioQuantumStatisticalTheory1971a,haakeQuantumStatisticsSuperradiant1972}. If, however, the initial state is superradiant, the radiation process is essentially classical and $g^{(2)}(0)$ approaches one in the large $N$ limit, even at $t=0$. This has recently been predicted and confirmed in cascaded superradiant cold-atom systems~\cite{bachEmergenceSecondOrderCoherence2026,tebbenjohannsPredictingCorrelationsSuperradiant2024,ferioli_emergence_2024}. We also note that $g^{(2)}(0)$ has been recognized as a parameter that determines whether or not a fully inverted, extended system decays in a superradiant burst~\cite{massonUniversalityDickeSuperradiance2022}.

If a fully inverted ensemble of emitters is placed inside a single-mode cavity, the dynamics show a train of hyperbolic secant pulses for the number of photons in the cavity mode~\cite{bonifacioCoherentSpontaneousEmission1970, keeling_quantum_2009}, where each pulse can be seen as an analogue to Dicke superradiance in free space. In such a case, which we call dynamical superradiance, it was shown theoretically that for a fully inverted state, $g^{(2)}(0)=2$, whereas $g^{(2)}(0)$ approaches one if there is some initial coherence~\cite{bonifacioCoherentSpontaneousEmission1970}, similar to Dicke superradiance.

If the many-emitter--cavity system is continuously pumped, one can enter the regime of steady state superradiance, also called a superradiant laser \cite{haake_superradiant_1993, meiserProspectsMillihertzLinewidthLaser2009, bohnetSteadystateSuperradiantLaser2012}. Note that in steady state superradiance, the second-order coherence is independent of the absolute time, such that $g^{(2)}(t,\tau)\equiv g^{(2)}(\tau)$. There exist three different regimes dependent on the strength of the incoherent pump $w$, the collective decay rate $\Gamma_c$ and the number of molecules $N$ \cite{meiserIntensityFluctuationsSteadystate2010,meiserSteadystateSuperradianceAlkalineearthmetal2010}: (i) Weak pumping $w\ll\Gamma_c$, which leads to photon bunching and $g^{(2)}(0)\geq2$, (ii) intermediate pumping $\Gamma_c<w<N\Gamma_c$ which is the superradiance regime, and $g^{(2)}(0)\approx1$ for large $N$ \cite{temnovPhotonStatisticsCooperative2009}, and (iii) strong pumping $w\gg N\Gamma_c$, which results in thermal radiation and hence $g^{(2)}(0)=2$.

Strong light–matter coupling in organic materials, enabled by the high binding energy of Frenkel excitons, has allowed polariton lasing and BEC phenomena at room temperature, with picosecond- or faster dynamics. 
Polariton emission from a strongly coupled plasmonic lattice, interpreted as BEC \cite{Vakevainen2020_our_condensate}, is an exciting example because it exhibits properties distinct from other strongly coupled systems. One difference is that the transition to condensation does not occur directly from active medium fluorescence; instead, the system passes through stages of polariton lasing and an intermediate regime, revealing a two-threshold dependence of emission intensity on pump fluence that clearly marks the transitions from fluorescence to lasing and from the intermediate regime to BEC. The system demonstrates a high degree of spatial coherence over distances of at least 0.5 mm and exceptionally short temporal coherence time ($<1$ ps) \cite{PhysRevLett.127.255301.Antti}, while also showing a ground state peak and high-energy photon distribution that fit the Bose-Einstein distribution at room temperature. The apparent combination of highly non-equilibrium character with essentially equilibrium concepts (BEC, thermalization) calls for further investigation. We pursue this by experimentally measuring the dynamics of the emission and its second-order coherence, and by theoretical calculations exploring the possibility of superradiance as an alternative interpretation to BEC with ultrafast thermalization.    

\section{\label{sec:main}Results}
\subsection{Experimental results}
\begin{figure}[ht!]
 \includegraphics[width=1\linewidth]{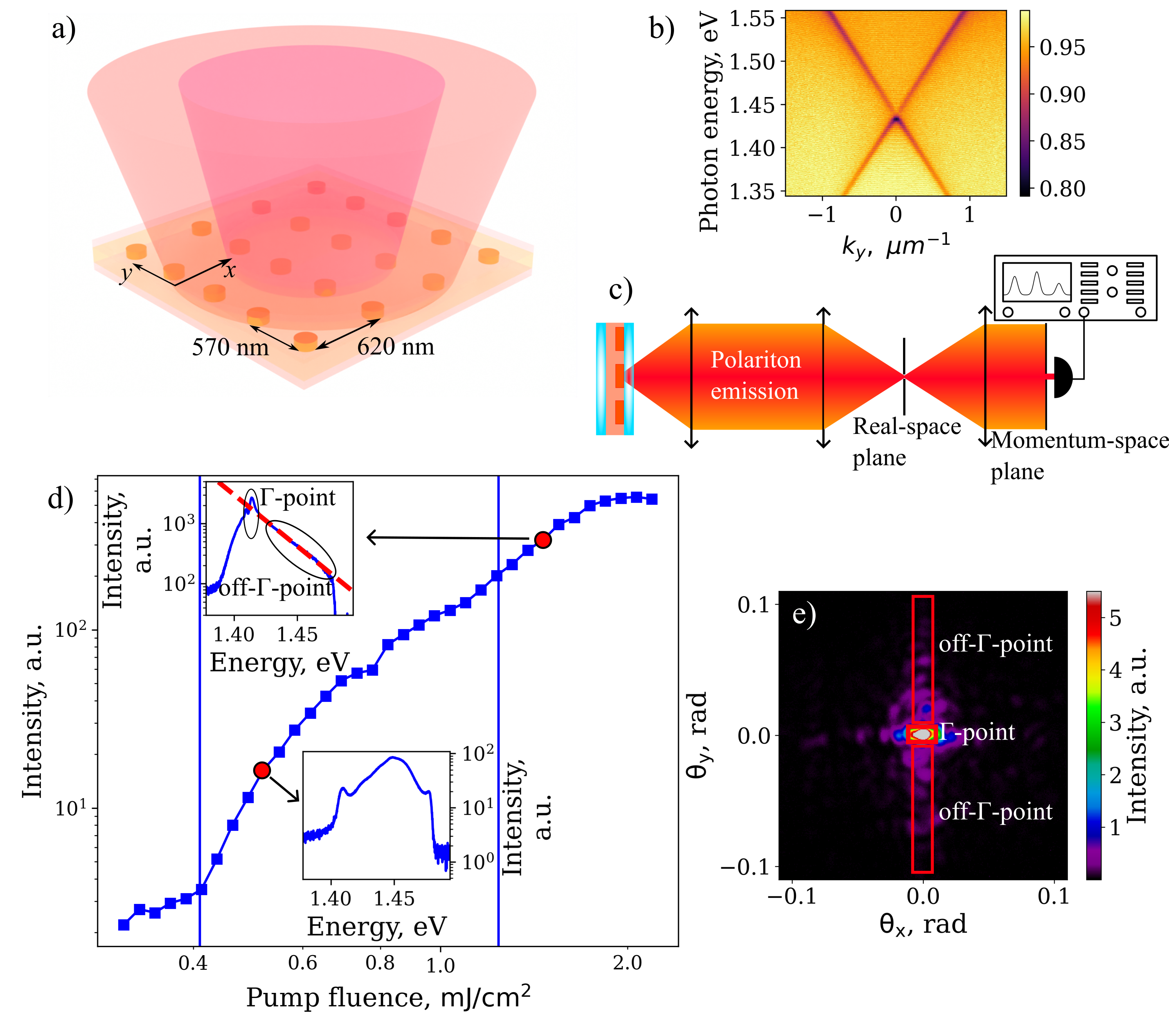}
 \caption{\label{fig1}a) Schematic of the studied structure with the coordinate frame and lattice periodicity indicated. b) Energy dispersion (transmission) of the structure as a function of the in-plane momentum component $k_y$. c) Scheme of the measurement of pulse-to-pulse fluctuation of the polariton emission. d) Dependence of the emission intensity on pump fluence. The two thresholds are denoted by vertical lines, and the characteristic spectra above the respective thresholds are shown in insets. e) Polariton emission momentum-space pattern with the areas corresponding to $\Gamma$-point and off-$\Gamma$-point states (see the inset of d)). The pump fluence is above the second threshold. The emission angles $\theta_x$ and $\theta_y$ correspond to inplane momenta $k_x$ and $k_y$ ranging from 0 to $\pm$1.57 $\mathrm{\mu m^{-1}}$.} 
 \end{figure}

To study polariton emission in a plasmonic lattice, we fabricated a rectangular array of metal nanoparticles identical to that used in Ref.~\cite{Vakevainen2020_our_condensate}, with lattice periods of $p_x = 620\,\mathrm{nm}$ and $p_y = 570\,\mathrm{nm}$ (Figure~\ref{fig1}.a). The optical modes are hybrids of lattice diffractive orders and the localized plasmonic resonances of the nanoparticles, called surface lattice resonances (SLRs) \cite{slr_rev1, slr_rev2}. They have dispersions with band edges that can serve as (ground) states for lasing or condensation (Figure~\ref{fig1}.b). As the active medium, IR-792 dye at a concentration of 80 mM was used to ensure the strong-coupling regime between the molecule absorption transition and the SLR modes. Further details on the sample fabrication and preparation are provided in the Methods section. The transverse-electric (TE)-polarized SLR mode relevant to the polariton emission studied here has its propagation direction along the $y$-direction; its $\Gamma$-point energy given by the $y$-period is 1.43 eV in the uncoupled regime. The mode associated with the other lattice periodicity has a $\Gamma$-point energy well outside the dye emission spectrum and does not contribute to the polariton emission. Pump radiation at 1.55 eV (800 nm), with 50 fs pulse duration and 1 kHz repetition rate, was used to excite polaritons. 

Several methods can be used to measure the second-order coherence function: Michelson interferometry with nonlinear-optical detection \cite{Boitier2009}, Hanbury Brown–Twiss (HBT) interferometry \cite{LAIHO2022128059}, and direct measurement of emission fluctuations (typically pulse-to-pulse) \cite{PhysRevLett.119.223603}. The last method has the simplest experimental implementation (Figure \ref{fig1}.c); however, it does not allow one to detect $g^{(2)}(\tau=0)$ (hereafter $g^{(2)}$) values below one and strictly requires single-mode detection both spatially and temporally; otherwise the measured value of the second-order coherence function is underestimated \cite{Sh.Iskhakov:12}. For the system under study, this approach proved the most suitable: dye stability issues make the use of a Michelson interferometer impractical, and HBT interferometry is unsuitable here because it does not provide temporal resolution due to the ultrafast nature of the polariton emission \cite{Vakevainen2020_our_condensate} and is mainly used for measurements of weak or nonclassical ($g^{(2)}<1)$ light \cite{loudon_quantum_2000}.

To measure pulse-to-pulse fluctuations, an amplified Si photodiode was used. The photodiode was placed in the Fourier plane of the sample, and a pinhole with a diameter of $250\,\mu\mathrm{m}$ was installed in front of the detector to ensure single-momentum-mode detection (see Methods section). The overall intensity dependence on the pump fluence shows a two-threshold profile typical for such systems \cite{Vakevainen2020_our_condensate}, see Figure~\ref{fig1}.d: bright $\Gamma$-point emission appears at the center, with a high-energy intensity distribution that depends on pump fluence. The latter corresponds to non-$\Gamma$-point TE-mode emission, see the momentum-space pattern for the polariton emission above the second threshold shown in Figure~\ref{fig1}.e. Moreover, due to the dispersion of the TE mode, each state in momentum space corresponds to its own energy. The photodiode was moved in the momentum-space plane to acquire separate measurements for the $\Gamma$- and non-$\Gamma$-point modes. The pump intensity increase is accompanied by the emission frequency blueshift (Figure \ref{fig:s_blueshift} for $\Gamma-$point).

 \begin{figure}[ht!]
 \includegraphics[width=1\linewidth]{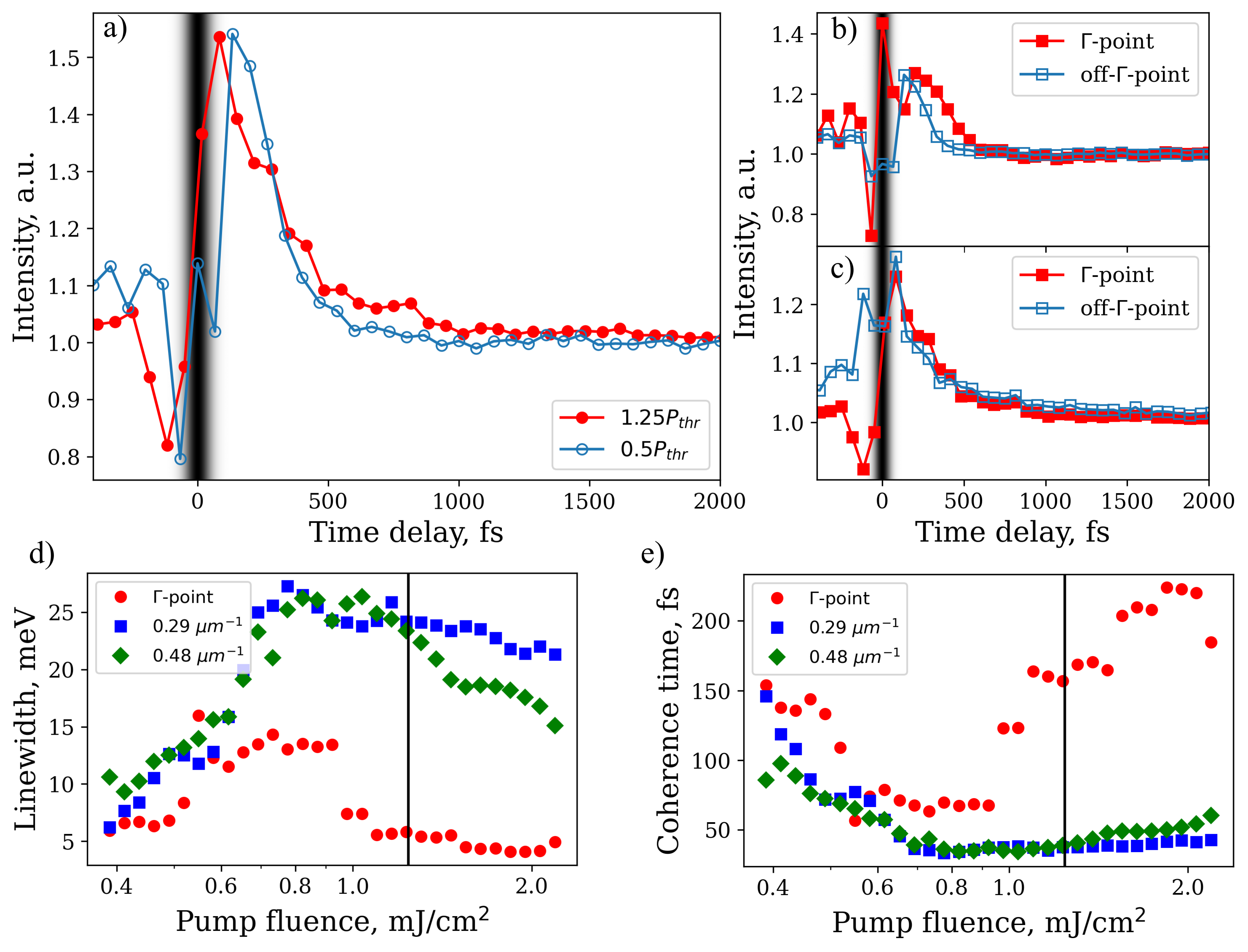}%
 \caption{\label{fig2}a)-c) Typical time profile of the polariton emission, black area denotes the pump and probe pulses' overlap. a) Time profiles of the total emission for pump fluence below and above the second threshold $P_\text{thr}$. Their profiles for $\Gamma$-point and non-$\Gamma$-point radiation are shown in panels b) (below the second threshold) and c) (above it). d) The emission linewidth for different states in momentum space. e) The corresponding coherence time within the assumption of Lorentzian line shape. The data in panels d) and e) are shown starting from the first threshold; the second threshold is denoted by a vertical line.} 
 \end{figure}

For the analysis of pulse-to-pulse fluctuations, a measurement of the polariton emission pulse duration is required, because approximate equality between the coherence time and the pulse duration is a prerequisite for correctly estimating $g^{(2)}$ in terms of the number of detected temporal modes. To this end, we used a technique analogous to double-pump spectroscopy, see e.g.~\cite{doi:10.1021/acs.nanolett.8b00531}: the sample was excited by two time-delayed pump pulses, one strong and one weak. The first pulse creates the main polariton population, while the second pulse, whose fluence is below that of the first pulse and below the polariton-lasing threshold, adds a small number of polaritons, which depends nonlinearly on the time delay and the polariton lifetime. During the experiment, momentum-space patterns were recorded, allowing the pulse durations of the $\Gamma$- and non-$\Gamma$-point emission to be determined separately. The results are shown in Figure~\ref{fig2}. The black-shaded area denotes the temporal overlap of the two pulses, leading to interference between them: an arbitrary increase or decrease in the total pump fluence and spatial modulation of the pump. Measurements were taken near the second threshold, both below and above it (Figure \ref{fig2}.a-c). In both cases, the emission pulses are short, with a full width at half maximum of $< 500$ fs. The overall shape of the curves resembles that of the temporal first-order coherence function in both shape and characteristic time \cite{PhysRevLett.127.255301.Antti}, supporting the conclusion that the $\Gamma$-point emission above the second threshold is fully coherent over the pulse duration. To estimate the coherence time for the non-$\Gamma$-point emission, and $\Gamma$-point emission below the second threshold, we fit the lineshape of the emission using a Lorentzian profile and then extracted the coherence time of the pulse (Figure \ref{fig2}.d,e; linewidths and coherence times for a broad range of $k_y$ are shown in Figure \ref{fig:s_linewidth}). At the lowest fluences, the linewidths reflect the pure SLR modes, which are largest at the $\Gamma$-point due to the largest plasmonic component, while higher energy modes are more light-like. These results, along with the pulse duration, show (see Methods section) that the values of $g^{(2)}(0)$ for $\Gamma$-point emission below the second threshold and for non-$\Gamma$-point emission are underestimated, but the underestimation of $g^{(2)}(0)-1$ is no more than a factor of 10 in these cases. In contrast, based on the above analysis, the measurement of $g^{(2)}(0)$ for the $\Gamma$-point emission above the second threshold is accurate.

\begin{figure}[ht!]
 \includegraphics{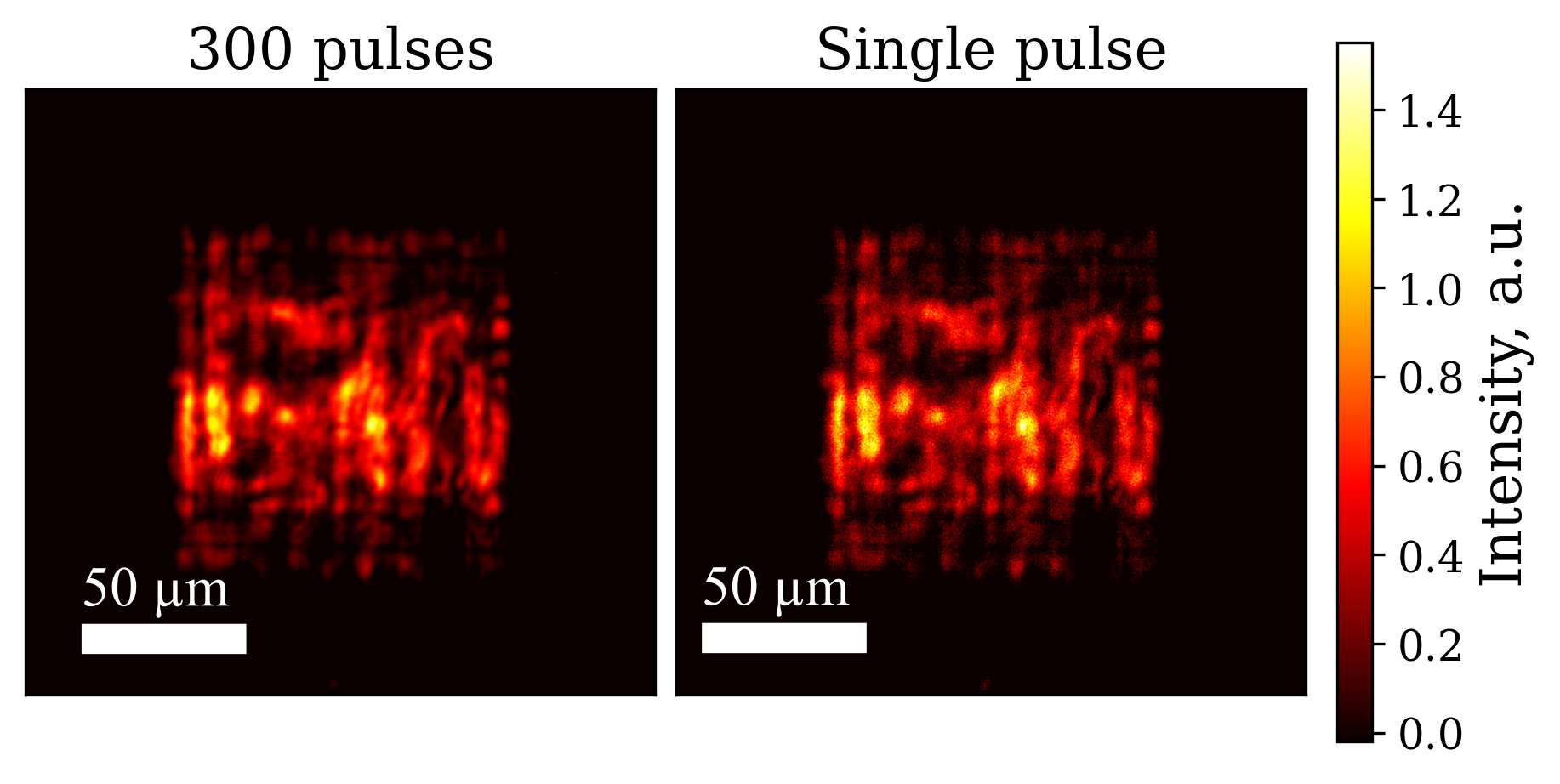}%
 \caption{\label{fig4}Averaged (left) and single-shot (right) real-space polariton emission patterns at the pump fluence 1.36 $\mathrm{mJ/cm^2}$ (1.1P$\mathrm{_{thr}}$, above the second threshold).}
 \end{figure}

Another criterion for the validity of the second-order coherence measurement approach that we use is that we consider a single mode. To maintain single-momentum-mode detection for measuring the second-order coherence function, two points in momentum space were selected: the $\Gamma$-point and one of the non-$\Gamma$-point states. Quite uniquely among organic polariton condensates, the real-space images of the polariton emission from the plasmonic lattice do not show any dynamical disorder (Figure~\ref{fig4}); any static fluctuations are caused by pump intensity inhomogeneity. Together with the stability of the momentum-space pattern (Figure \ref{fig:s_rs}), this indicates the stability of both the intensity and phase patterns of the emission \cite{bec_jani}. This simplifies the interpretation of our measurements, as we do not need to consider dynamic disorder effects.

\begin{figure}[ht!]
 \includegraphics[width=0.9\linewidth]{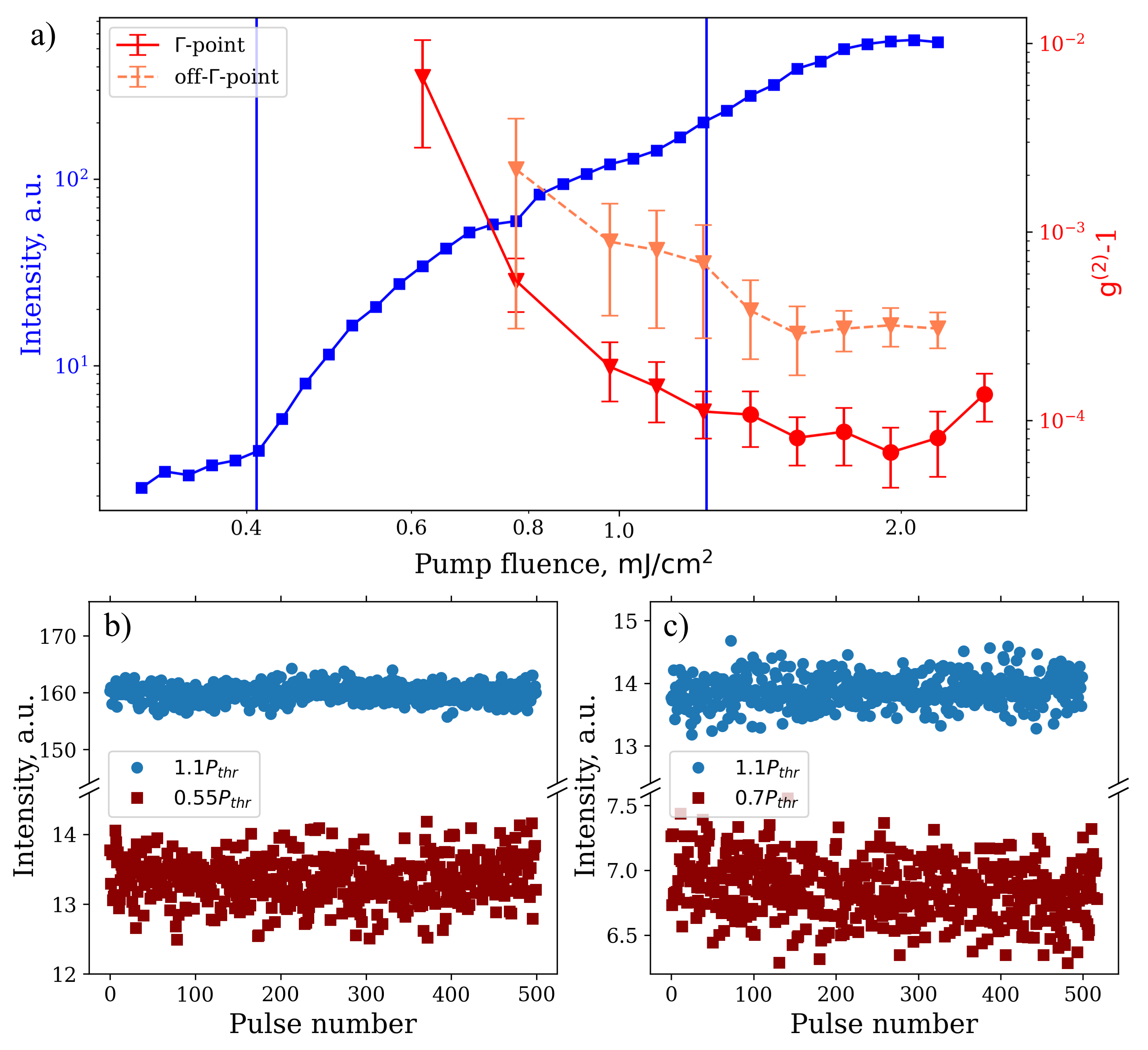}%
 \caption{\label{fig3}a) The $g^{(2)}(0)$ dependence on pump fluence for $\Gamma$-point (red curve) and non-$\Gamma$-point (orange curve) states. As a reference, the dependence of polariton emission intensity on pump fluence is shown; the two vertical lines denote the first and second thresholds. The triangles denote underestimated values of $g^{(2)}(0)$. b) and c) show fluctuation suppression for $\Gamma$-point and off-$\Gamma$-point, respectively, for pump fluences below and above the second threshold.}
 \end{figure}

As a reference for the second-order coherence measurements, the $g^{(2)}$ of the pump laser was measured, yielding $g^{(2)} - 1 = 1.2 \times 10^{-5}$. This small deviation from one, expected for a coherent laser, is attributed to the electronic noise of the detection system rather than laser intensity fluctuations, and can be considered as the minimum measurable value of $g^{(2)}$. Measurements of the second-order coherence function for the plasmonic lattice samples were then performed at different pump fluences. The resulting curves, along with the emission intensity versus pump fluence for reference, are shown in Figure \ref{fig3}.a. The two thresholds are clearly visible and indicated by the vertical lines. Near the first threshold, from fluorescence to polariton lasing, the emission intensity was too low to record single-shot signals, but the second-order coherence is expected to start at two and then decrease toward one. Between the first and second thresholds, $g^{(2)}$ for both $\Gamma$-point and non-$\Gamma$-point emission remains close to one, even when the underestimation of the $g^{(2)}$ values is taken into account. At the second threshold both curves reach their lowest values: $g^{(2)} - 1 \approx 8 \times 10^{-5}$ for the $\Gamma$-point and $g^{(2)} - 1 \approx 3 \times 10^{-4}$ for the non-$\Gamma$-point state, with $g^{(2)}_{\Gamma\text{-point}} < g^{(2)}_{\text{non-}\Gamma\text{-point}}$ for all measured pump fluences. At higher pump fluences, $g^{(2)}$ for the $\Gamma$-point increases due to system degradation. Notably, the $g^{(2)}$ values for the $\Gamma$-point radiation above the second threshold are closer to one than the $g^{(2)}$ measured for the total number of polaritons in a BEC system with negative detuning at a much higher pump fluence ($5\,P_{\mathrm{thr}}$) \cite{PhysRevB.Anton}. Typical pulse-to-pulse fluctuations of the polariton emission are shown in Figure~\ref{fig3}.b,c for pump fluences below and above the second threshold, respectively. Both traces show no evidence of sample emission degradation, and a clear suppression of fluctuations is observed above the second threshold.

\begin{figure}
    \centering
    \includegraphics[width=1\linewidth]{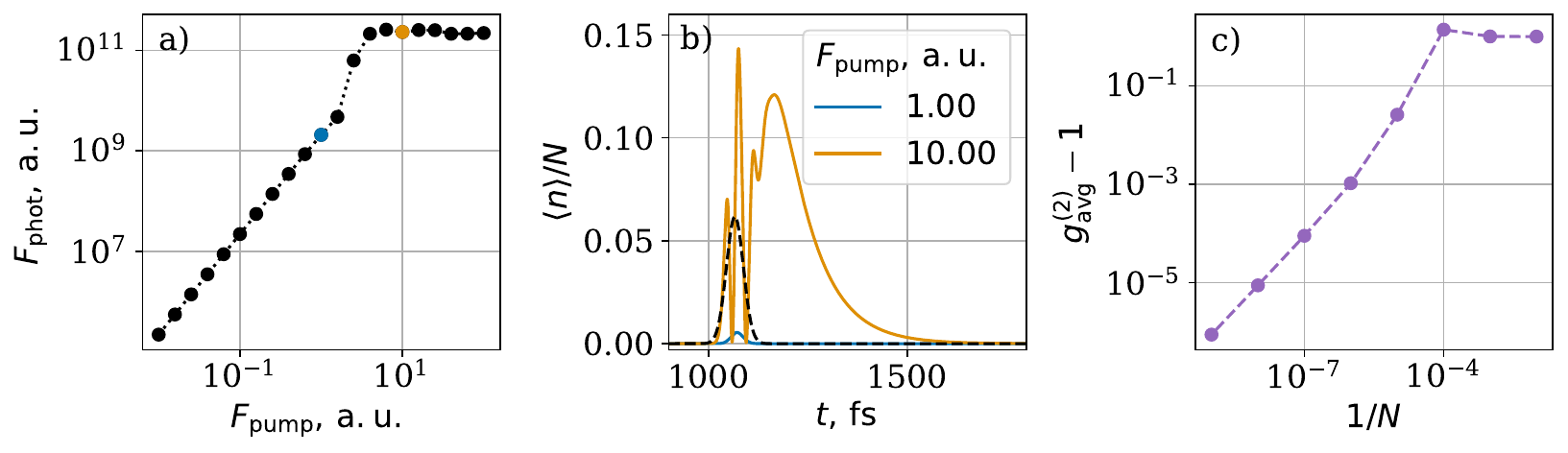}
    \caption{Numerical simulations of the photon emission. a) Photon fluence as a function of pump fluence for a pump pulse with $\tau_\text{FWHM}=50\fs$. b) Average number of photons in the cavity mode as a function of time for two different pump fluences, indicated by the colored data points in a). The black dashed line shows the shape of the pump pulse. c) Second-order coherence calculated from the Holstein-Tavis-Cummings model starting from an initial state with non-zero electronic coherence for varying numbers of (nearly) excited molecules and zero pump. All other parameters not specified are listed in Table~\ref{tab:parameters}.}
    \label{fig:threshold_n}
\end{figure}

\subsection{Theoretical model}
Theoretical description of an organic polariton BEC is difficult due to the system's highly non-equilibrium nature and presence of the vibrational degrees of freedom in organic molecules.
State-of-the-art models treat the Frenkel excitons as two-level systems, which are vibrationally dressed due to the coupling to the molecular phonons, and also couple to the cavity mode(s) in a Tavis-Cummings-type Hamiltonian. Incoherent processes such as pumping, exciton- and photon losses, and exciton dephasing can be included in a Lindblad master equation, which usually cannot be solved exactly for a macroscopic number of molecules due to the exponentially growing Hilbert space. Thus, one must resort to some approximations such as mean-field~\cite{strashkoOrganicPolaritonLasing2018} and second-order cumulant ~\cite{arnardottirMultimodeOrganicPolariton2020} approximations, or a fast thermalization approximation~\cite{shishkovExactAnalyticalSolution2022}. Moreover, kinematic rate equations for the populations of the polaritons have been used successfully to model mainly inorganic polariton condensation~\cite{tassoneBottleneckEffectsRelaxation1997,malpuechRoomtemperaturePolaritonLasers2002,caoCondensationKineticsCavity2004,doanMicrocavityPolaritonKinetics2006,doanCoherenceCondensedMicrocavity2008}, but can also be used for organic polaritons~\cite{ishiiKineticModelsPolariton2025}.
For a more detailed review of theoretical approaches specifically for organic polariton condensation, see Ref.~\cite{keeling_boseeinstein_2020}. 

We now propose an alternative explanation for the experimental results based on superradiance in vibrationally dressed molecules. Superradiance describes the coherent spontaneous decay of an ensemble of molecules, resulting in short, high-intensity emission pulses. In our theoretical approach, we place the molecules in a cavity, resulting in a model for dynamical superradiance~\cite{freterTheoryDynamicalSuperradiance2025}. Temperature enters via occupation of the vibrational excitations and in their thermalization rate. We limit the description to one optical mode, and we include one (dominant) vibrational mode of the molecules explicitly in the system Hamiltonian. This setup allows a straightforward application of the \emph{QuantumCumulants.jl} package~\cite{plankensteinerQuantumCumulantsjlJuliaFramework2022} for deriving mean-field and second-order cumulant equations, to numerically calculate the dynamics for a macroscopic number of molecules. Limitations of this approach are the neglect of multiple cavity modes involved in the polariton dynamics, and of all but one vibrational mode in the molecules. 

\begin{table}[ht]
    \centering
    \begin{tabular}{|c|c|c|}
        \hline
        Parameter & Explanation & Typical values \\ \hline
        $\wc$ & Frequency of cavity mode & $1.43\eV$  \\
        $\omega_0$  & Frequency of the two-level system &  $1.53$~eV (IR-792)\\
        $\omega_L$ &Frequency of laser drive & $\sim\omega_0$\\
        $\omega_\nu$ &Frequency of molecular vibration & $0.15$~eV \cite[Figure 1]{heilmannStrongCouplingOrganic2020}\\
        $g\sqrt{N}$ & Collective light-matter coupling & $0.1\eV$ (Figure~\ref{fig:s_strong})\\
        $N$ & Number of molecules & $\sim10^{10}$\\
        $\kappa$  & Cavity loss of photons & $0.01\eV$ (Figure~\ref{fig:s_strong}) \\
        $\gamma$ & Independent decay & $1/\gamma=1\,\text{ns}\Rightarrow \gamma=6.58\times10^{-7}\eV$ \\
        $\gamma_\phi$ & Individual dephasing & $0.025\eV$ \\
        $\gamma_\nu$ &Thermalization rate of vibrations & $\sim0.05\eV$\\
        $S$ &Huang-Rhys factor & $~\sim 0.14$ \\
        $T$& Temperature of vibrations & $0.026\eV$\\
        \hline
    \end{tabular}
    \caption{Typical values of parameters used in the numerical simulations.}
    \label{tab:parameters}
\end{table}


Specifically, we treat the ensemble of $N$ molecules as identical two-level systems described by Pauli operators $\sigma_i^{\pm,z}$ with transition frequency $\omega_0$ between the electronic states, where the index $i$ labels the molecule. The electronic states couple to a vibrational mode with frequency $\omega_\nu$ and bosonic annihilation operator $b_i$, as well as to a cavity mode with frequency $\wc$ and bosonic annihilation operator $a$. In the rotating-wave approximation, this results in the Holstein-Tavis-Cummings model~\cite{cwikPolaritonCondensationSaturable2014, spanoOpticalMicrocavitiesEnhance2015, herreraCavityControlledChemistryMolecular2016, wuWhenPolaronsMeet2016}. Additionally, we include a coherent pump with envelope $\xi(t)$ and frequency $\omega_L$, giving the Hamiltonian ($\hbar=1$)
\begin{multline}
        H = \tilde\omega_\mathrm{c} a^\dagger a + \sum_{i=1}^N\left[\frac{\tilde\omega_0}{2}\sigma_i^z + g(a\sigma_i^+ + a^\dagger \sigma_i^-) + \omega_\nu b^\dagger_i b_i + \omega_\nu\sqrt{S}(b^\dagger_i + b_i)\sigma_i^z\right]
        +\sum_{i=1}^N \xi(t)(\sigma_i^+
        +\sigma_i^-),\label{eq:H}
\end{multline}
where we moved into a rotating frame with frequency $\omega_L$, such that $\tilde{\omega}_\mathrm{c}=\wc-\omega_L$ and $\tilde\omega_0=\omega_0-\omega_L$.
The model contains two coupling parameters: the individual light-matter coupling $g$, which is related to the Rabi-splitting $\Omega=2g\sqrt{N}$, and the dimensionless Huang-Rhys factor $S$, which determines the coupling strength of the electronic degrees of freedom to the vibrational mode. 
We employ a Gorini-Kossakowski-Sudarshan-Lindblad~\cite{gorini_completely_1976, lindblad_generators_1976} master equation to include cavity losses at rate $\kappa$, individual molecular losses at rate $\gamma$, individual dephasing at rate $\gamma_\phi$, and vibrational thermalization at rate $\gamma_\nu$:
\begin{equation}
    \partial_t\rho = -\im[H+H_\text{LS},\rho] + \kappa\mathcal{L}[a] + \sum_{i=1}^N\left(\gamma\mathcal{L}[\sigma_i^-] + \gamma_\phi\mathcal{L}[\sigma_i^z] + \gamma_\nu n_B\mathcal{L}[b_i^\dagger] + \gamma_\nu(n_B+1) \mathcal{L}[b_i]\right),\label{eq:master1}
\end{equation}
where $H_\text{LS}=\im\frac{\gamma_\nu}{4}\sum_{i=1}^N(b^\dagger_ib^\dagger_i  - b_ib_i)$ is a Lamb-shift term~\cite{mannouch_ultra-fast_2018}, $\mathcal L[X] = X\rho X^\dagger - (X^\dagger X\rho + \rho X^\dagger X)/2$ and $n_B= (\exp(\omega_\nu/T) - 1)^{-1}$ is the Bose-Einstein distribution at temperature $T$ ($k_B=1$). 

Except for the coherent pump, the model described above is equivalent to the model used in Ref.~\cite{freterTheoryDynamicalSuperradiance2025} to describe dynamical superradiance in organic molecules starting from an (almost) fully inverted initial state. In contrast, we choose here an initial state with zero photons in the cavity mode, the two-level systems in the ground state, and the vibrational modes in a thermal state. The system is then driven by the coherent pump. For further details on the theoretical model, see Supplementary Material~B.

We use second-order cumulant and mean-field approximations to integrate Eq.~\eqref{eq:master1} numerically. Typical values for the parameters used in the simulations are given in Table~\ref{tab:parameters}. All values are realistic and correspond to our experimental system, except for individual dephasing, thermalization rate of vibrations, and Huang-Rhys factor, which are obtained by fitting the absorption spectrum of the Holstein model for a single molecule (Eqs.~\eqref{eq:H} and~\eqref{eq:master1} without the cavity mode and coherent pumping) to an experimentally measured absorption spectrum. In Figure~\ref{fig:threshold_n}(a), we plot the photon fluence defined by $F_\text{phot}=\int\langle a^\dagger(t)a(t)\rangle\dd t$ as a function of the pump fluence $F_\text{pump} = \int \xi(t)\dd t$, showing a clear threshold behavior. In Fig~\ref{fig:threshold_n}(b), the number of photons in the cavity mode $\langle n\rangle/N$ as a function of time, below and above threshold, is shown. Above threshold, the photon number is appreciably larger than below threshold and shows oscillations corresponding to the detuning $|\Delta|=|\wc-\omega_0|$.

Next, we calculate an estimate for the second-order coherence function $g^{(2)}(0)$ as a function of the number of molecules in the system, using a second-order cumulant approach. In this calculation, we eliminate the coherent pump for simplicity and replace it with initial conditions resembling an almost fully inverted ensemble of molecules with an initial coherence inherited by the pump. For details on the $g^{(2)}(0)$ calculations, see Supplementary Material~B. The results in Figure~\ref{fig:threshold_n}(c) show that $g^{(2)}(0)$ approaches one for a macroscopic number of molecules. We also checked that the calculated $g^{(2)}(0)$ values obtained using second-order cumulants converge to the exact value obtained using quantum trajectories in the case of zero vibrational coupling (see Figure~\ref{fig:g2_c2_exact}).

Since our current model is restricted to a single cavity mode, it is not possible to obtain the Bose-Einstein distribution of population at different modes above the second threshold. Since temperature enters the theoretical calculations via the molecular vibrational level occupation factors, it is possible that a multilevel description of the molecules could describe a thermally distributed high-energy tail as in the Bose-Einstein distribution. Theoretical studies of superradiance in multilevel systems suggest that emission would occur on multiple transitions, although in the large $N$ limit, the transition with the largest rate dominates~\cite{holzingerSymbolicQuantumTrajectoryMethod2025}. Our present theoretical results, although limited to a single mode, show that the experimentally observed second-order coherence value close to one and the extremely fast time scales of the emitted pulses are consistent with the superradiance picture. 

\section{Conclusions}

We presented a detailed experimental and theoretical study of second-order coherence in ultrafast polariton emission from strongly coupled plasmonic lattices. We measured $g^{(2)}(0)$ on the relevant sub-picosecond timescale by resolving pulse-to-pulse fluctuations, and determined it separately for the $\Gamma$-point condensate-like emission and the non-$\Gamma$ high-energy tail. The emission pulses were found to be shorter than 500 fs, confirming dynamics in the ultrafast, THz regime, and supporting the validity of the coherence analysis. Remarkably, the measured coherence is extremely high: near the second threshold, $g^{(2)}(0)-1$ reaches approximately $8\times10^{-5}$ for the $\Gamma$-point emission. The interpretation is strengthened by careful control of single-mode detection as well as by the observed dynamical stability of the samples: single-shot and averaged real- and momentum-space patterns show no appreciable dynamical disorder. We showed that a vibrationally dressed Holstein–Tavis–Cummings model can reproduce threshold behavior and ultrafast emission, as well as $g^{(2)}(0)\to1$ for macroscopic ensembles. We propose that the experimental observations, while consistent with polariton BEC formed by sub-50-fs thermalization, may alternatively be understood as superradiance in a system hosting vibrational degrees of freedom. This interpretation broadens the discussion of coherence in driven light-matter condensates and may have implications for other reported condensates where high coherence, ultrafast dynamics, and collective emission coexist.

\section{Methods}
\subsection{Sample}
The gold nanoparticle array was fabricated with electron beam lithography on a borosilicate substrate (n = 1.52). A 2~nm layer of titanium was used for better adhesion of the gold nanoparticles to the substrate. The size of the array was 100x100 $\mathrm{\mu m}$. The array had a periodicity of 570~nm in the y-direction and 620~nm in the x-direction (Figure \ref{fig1}), providing TE-polarized SLR mode with the $\Gamma-$point energy within the dye emission range in y-direction. The individual nanoparticle had a cylindrical shape with a diameter of 100~nm and a height of 50~nm.  For polariton emission experiments, a dye solution of IR-792 (concentration of 80~mM) in a 2:1 mixture of benzyl alcohol and dimethyl sulfoxide was prepared; the refractive index of the dye solution matches that of the substrate. Then, a cavity formed by the substrate, another borosilicate slide, and a press-to-seal silicone ring (0.8~mm thickness) was filled with the solution. This particular sample preparation provides strong light-matter coupling with a Rabi splitting of 200~meV (Figure \ref{fig:s_strong}). For dispersion measurements, the samples were immersed in index-matching oil (dispersion of uncoupled system) or IR-792 dye solution (dispersion of coupled system) and sealed with a cover slip.

\subsection{Lasing and transmission measurement setup}
The scheme of the experimental setup is shown in Figure \ref{fig:s_setup}. As a pump beam for the polariton emission experiments, the horizontally (or x-) polarized output of the Coherent Astrella laser was used with the following parameters: central wavelength 800 nm (1.55 eV), pulse duration 50 fs, pulse repetition rate 1 kHz. For the dispersion (transmission) measurements, a white light source (halogen lamp) was used. The emission from the sample was collected with a 0.3/10x Nikon microscope objective, corresponding to the collection angle of $\mathrm{\pm17.5^{\circ}}$ with respect to the normal incidence to the sample. The back focal plane of the objective is imaged onto the spectrometer slit, providing both spectral and angular resolution in the y-direction (according to the sample coordinate frame, Figure \ref{fig1}) of the signal detected by the spectrometer CCD sensor. The sample anisotropy (rectangular lattice) ensures that the TE-polarized SLR-assisted polariton emission under study occurs only along the y-direction.  Two additional CMOS cameras were used to capture real- and momentum-space images of the emission. A longpass filter with a cut-off wavelength of 834 nm was used to filter out pump radiation in the detection channel. In the polariton emission experiments, the sample was pumped through the same objective that was used for the emission detection.

\subsection{Second-order coherence measurements}

For the second-order coherence measurements, the photodiode Thorlabs PDA10A2, placed in the momentum-space plane, was used. A pinhole with a diameter of 250 $\mathrm{\mu m}$ was installed in front of the photodetector. Taking into account the focal length of the tube lens and the momentum-space lens of 200 mm and 250 mm, respectively, this pinhole size provides single-momentum-mode detection ($\Delta k \Delta x \approx 2\pi$, where $\Delta x$ is the array size, $\Delta k$ is the size of momentum space cropped by the pinhole).

In case of multimode detection, experimental values of the second-order coherence function $g_{\text{exp}}^{(2)}(0)$ are underestimated in the following way \cite{Sh.Iskhakov:12}:
\begin{equation}
    g_{\text{exp}}^{(2)}(0) = 1+\frac{g_{\text{real}}^{(2)}(0)-1}{N},
\end{equation}
where $N$ is the number of modes. If we consider more specifically only temporal modes, in terms of pulse emission time $\tau_p$ and coherence time $\tau_c$, the formula can be rewritten as \cite{PhysRevB.Anton}:
\begin{equation}
    g_{\text{exp}}^{(2)}(0) = 1+\frac{g_{\text{real}}^{(2)}(0)-1}{\tau_p/\tau_c}.
\end{equation}

\section{\label{sec:ack}Acknowledgments}
We are thankful to Denis Kopylov, Vladislav Yu. Shishkov, Anton Zasedatelev, and Jonathan Keeling for fruitful discussions, and to Pavel Kliuiev for initial experiments on measuring second-order coherence with a different scheme.

The work was supported by the Jane and Aatos Erkko Foundation and the Technology Industries of Finland Centennial Foundation as part of the Future Makers funding program, by the Research Council of Finland under project number 349313, and by the Research Council of Finland through the Finnish Quantum Flagship project 358877. The work is part of the Research Council of Finland Flagship Programme, Photonics Research and Innovation (PREIN), decision number 346529, Aalto University. This work is part of the Finnish Centre of Excellence in Quantum Materials (QMAT). 

Part of the research was performed at the OtaNano Nanofab cleanroom (Micronova Nanofabrication Centre), supported by Aalto University.

\section{Author contribution}
P.T. and E.A.M. conceived the project. P.T. supervised the project. E.A.M. and S.E. fabricated the samples. E.A.M. built the photon statistics measurement setup and did the measurements. S.E. helped with the measurements. L.F. made the numerical simulations. E.W.L. contributed to the simulations. E.A.M., L.F., and P.T. wrote the manuscript, with contributions from other authors. All authors discussed the results and reviewed the manuscript. 

\section{Competing interests}
The authors declare that they have no conflict of interest.

\section{Data Availability Statement}
The data that support the findings within this manuscript are available from the corresponding author upon reasonable request.


\clearpage
\appendix


\title{\textbf{Supplementary Materials: Second-order coherence properties of ultrafast polariton dynamics in plasmonic lattices} 
}%

\author{Evgeny A.~Mamonov}
\author{Lukas Freter}
\author{Sioneh Eyvazi}
\affiliation{%
Department of Applied Physics, Aalto University School of Science, Aalto FI-00076, Finland
}%
\author{Elliot W.~Lloyd}
\affiliation{SUPA, School of Physics and Astronomy, University of St Andrews, St Andrews, KY16 9SS, UK}
\author{Päivi Törmä}%
 \email{Contact author: paivi.torma@aalto.fi}
\affiliation{%
Department of Applied Physics, Aalto University School of Science, Aalto FI-00076, Finland
}%

\maketitle

\setcounter{equation}{0}
\setcounter{figure}{0}
\setcounter{table}{0}
\setcounter{page}{1}
\makeatletter
\renewcommand{\theequation}{S\arabic{equation}}
\renewcommand{\thefigure}{S\arabic{figure}}

\subsection{Supplementary figures}
\begin{figure}[ht!]
    \centering
    \includegraphics[width=0.6\linewidth]{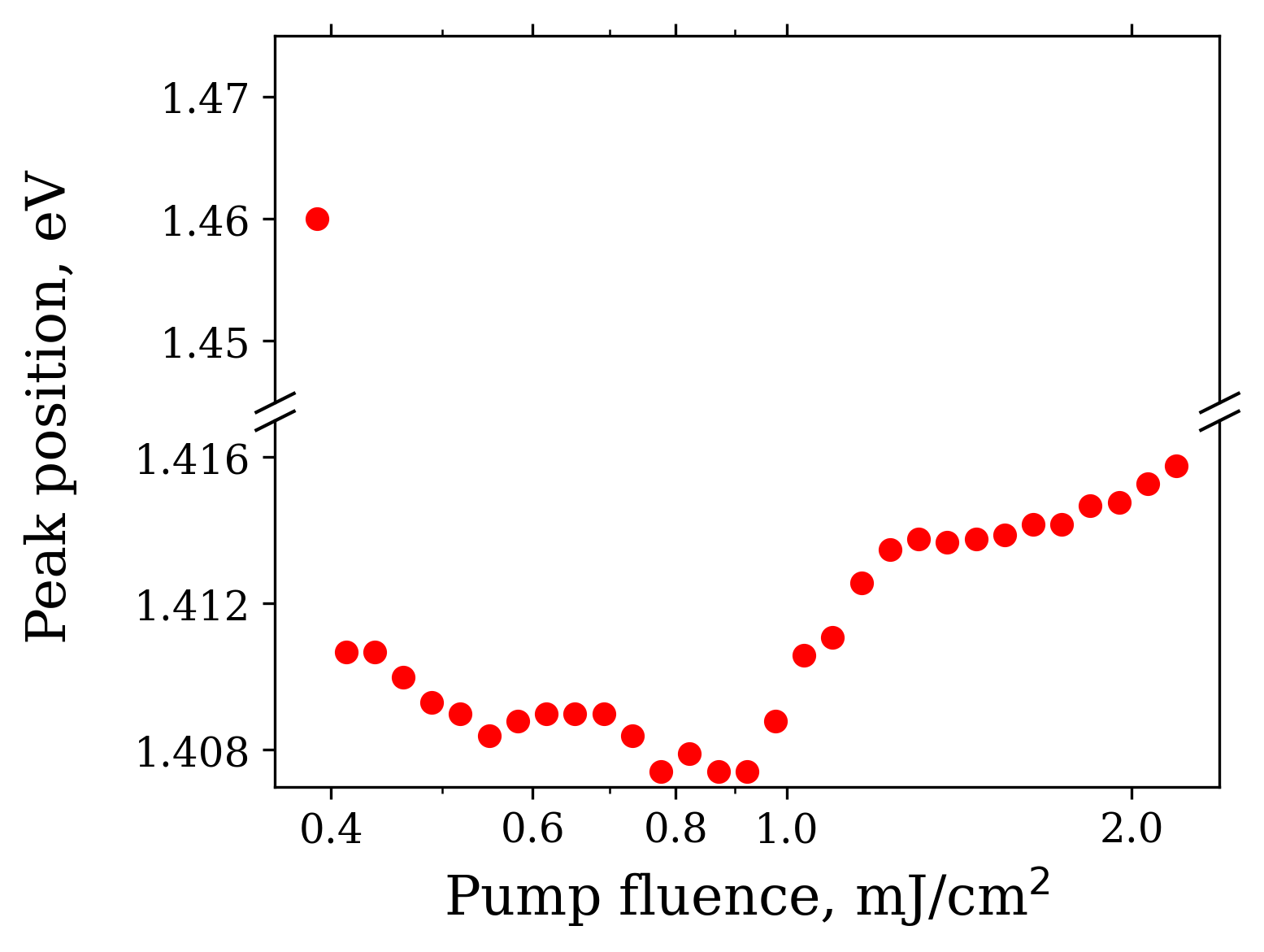}
    \caption{The spectral position of the polariton emission peak at $\Gamma$-point.}
    \label{fig:s_blueshift}
\end{figure}

\begin{figure}[ht!]
    \centering
    \includegraphics[width=1\linewidth]{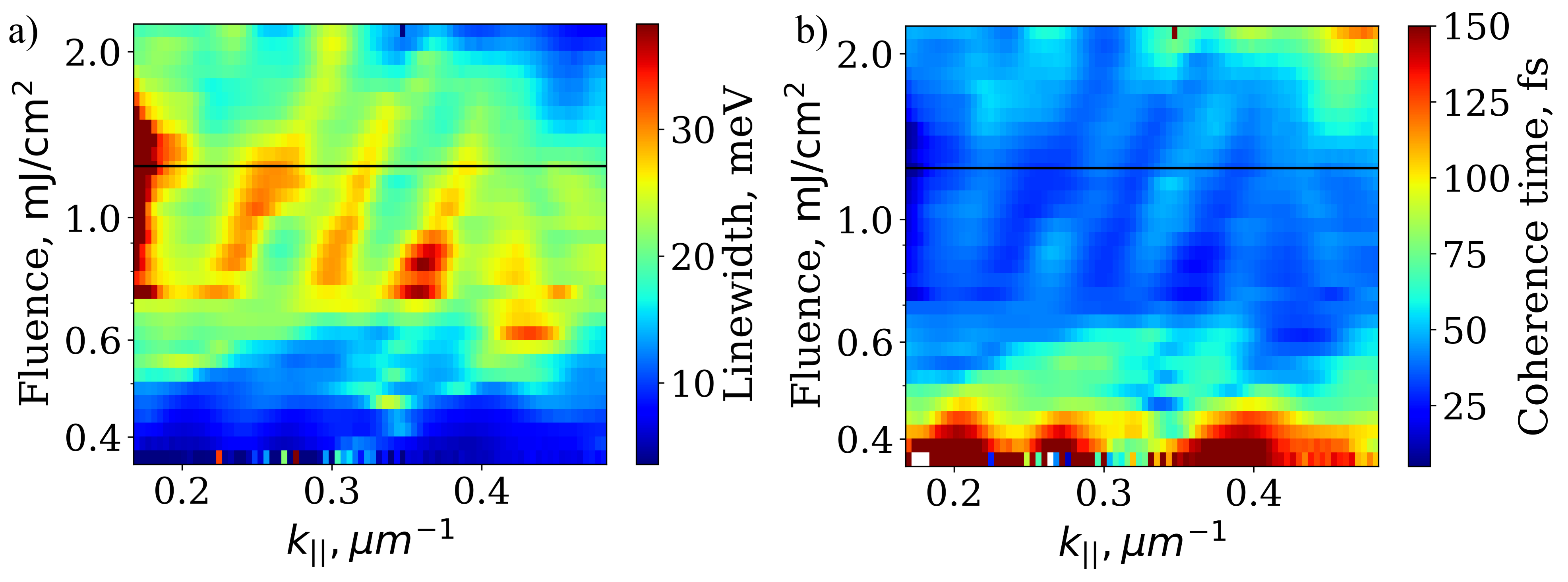}
    \caption{The linewidth and coherence time within the assumption of Lorentzian line shape for off-$\Gamma$-point states.}
    \label{fig:s_linewidth}
\end{figure}

\begin{figure}[ht!]
    \centering
    \includegraphics[width=0.9\linewidth]{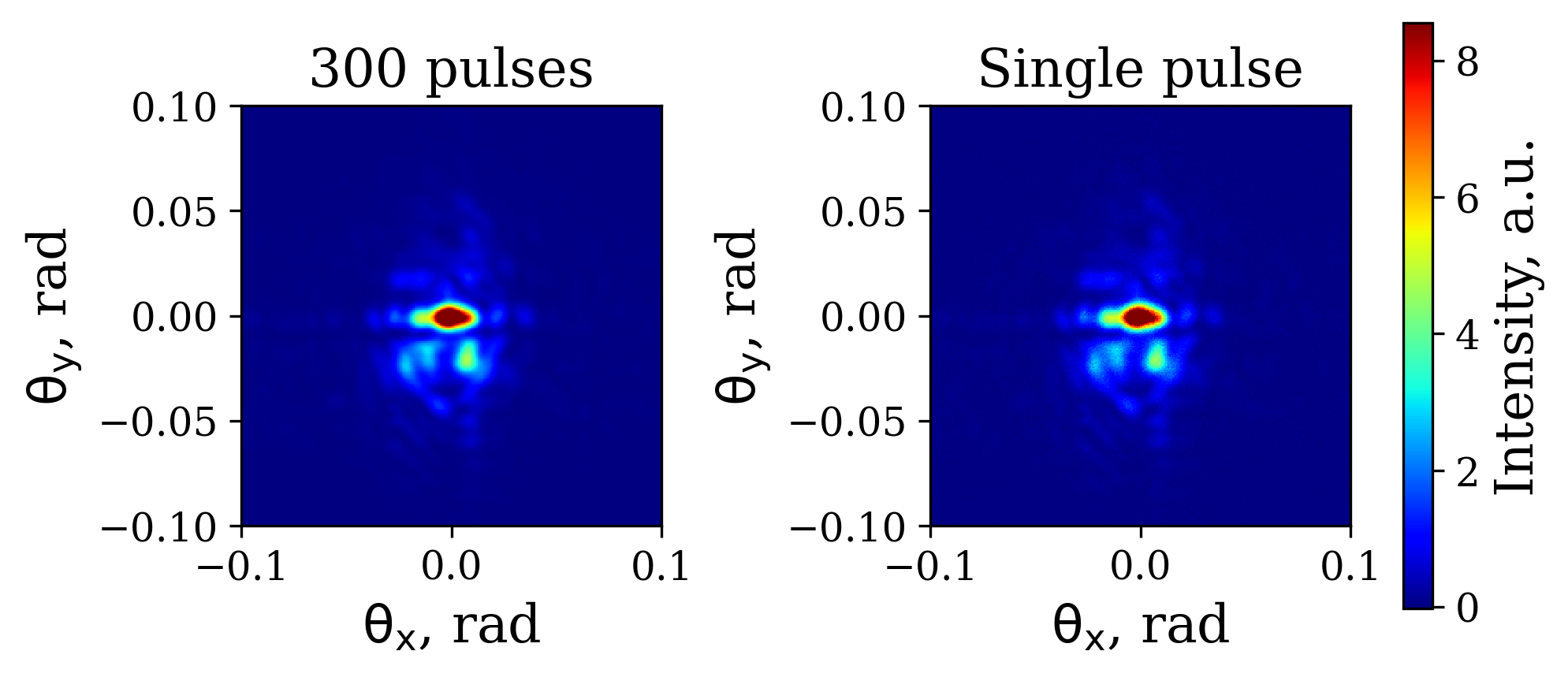}
    \caption{Averaged (left) and single-shot (right) momentum-space polariton emission patterns at the pump fluence 1.36 $\mathrm{mJ/cm^2}$ (1.1P$\mathrm{_{thr}}$, above the second threshold).}
    \label{fig:s_rs}
\end{figure}

\begin{figure}[ht!]
    \centering
    \includegraphics[width=0.9\linewidth]{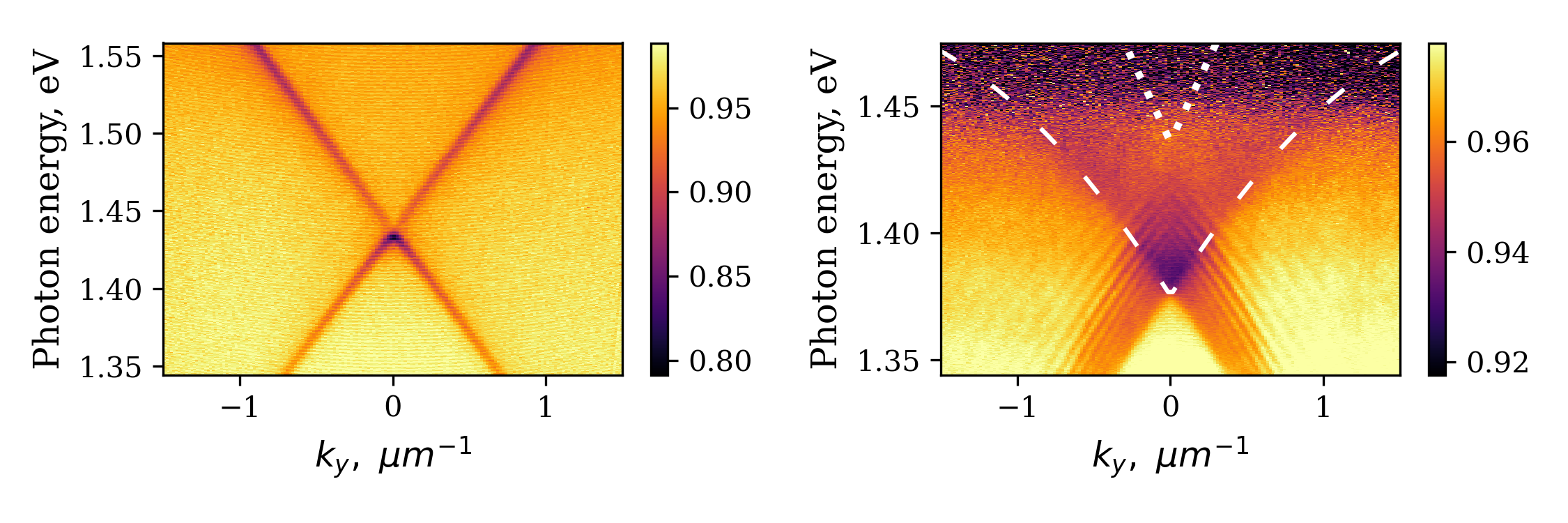}
    \caption{Dispersion (transmission) of the plasmonic lattice under study, without molecules on the left, and with them on the right. The dotted line denotes the dispersion of the uncoupled system, and the dashed line denotes the dispersion of the strongly coupled system obtained from a coupled-mode-theory fit, giving a Rabi splitting of $\Omega = 200$meV, which is larger than the average of the linewidths of the SLR mode and the absorption spectrum, namely 10 meV and 120 meV, respectively.}
    \label{fig:s_strong}
\end{figure}

\begin{figure}[ht!]
    \centering
    \includegraphics[width=1\linewidth]{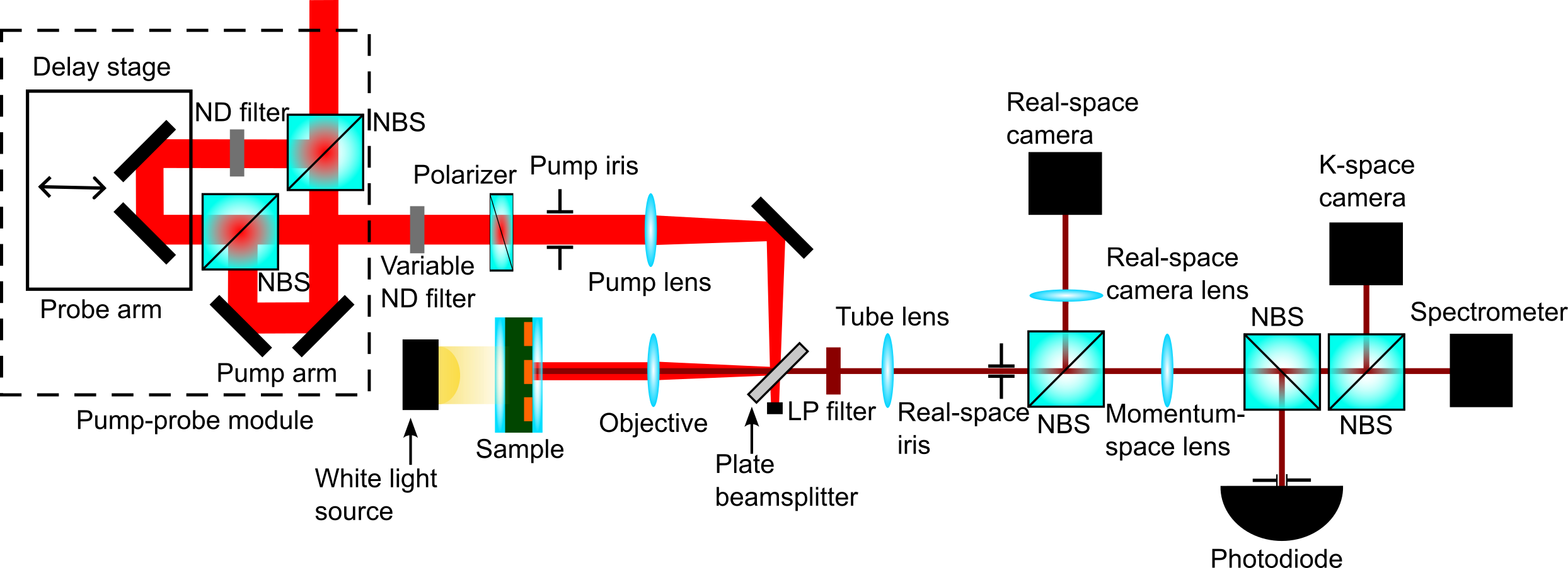}
    \caption{The full scheme of the experimental setup. NBS stands for non-polarizing beamsplitter, ND for neutral density, and LP for long-pass filter.}
    \label{fig:s_setup}
\end{figure}

\begin{figure}
    \centering
    \includegraphics[width=0.5\linewidth]{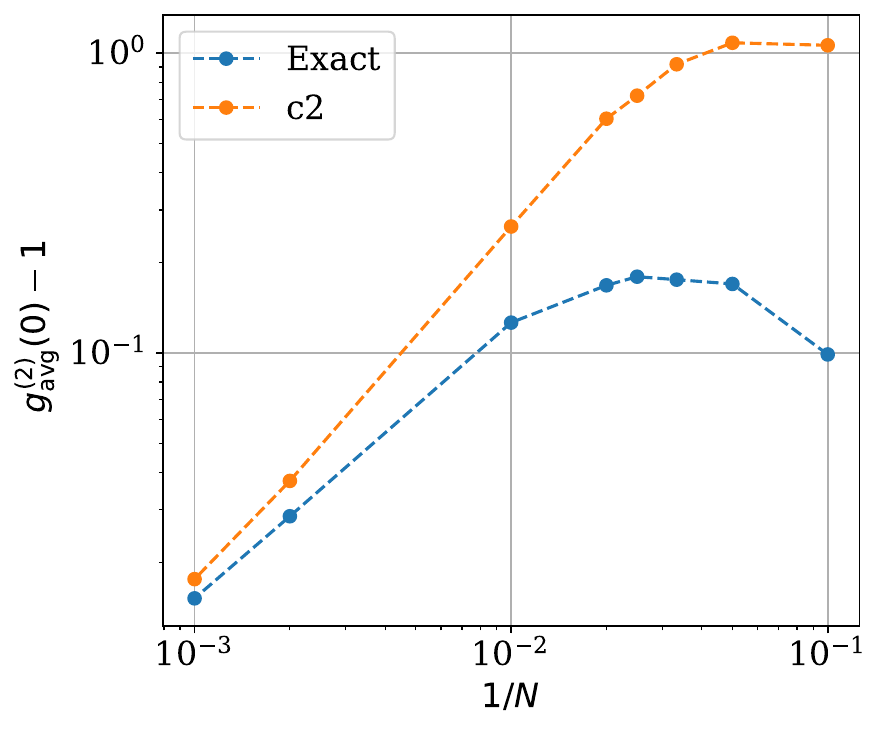}
    \caption{Calculated average second-order coherence as a function of the number of molecules $N$ for the model described in Eqs.~\eqref{eq:H},\eqref{eq:master1} in the main text with $S=0$ and no coherent pump. The blue curve is obtained from an exact quantum trajectories method, and the orange curve from second-order cumulants. The value of $g^{(2)}$ was averaged from $t=0$ until the first local minimum of the photon number, as described in Eq.~\eqref{eq:g2_avg}. Parameters: $g\sqrt{N}=0.2\eV, \Delta=\omega_\mathrm{c}-\omega_0=0,\,\kappa=0.01,\,\gamma=0,\,\gamma_\phi=0.01\eV,\,\theta=0.1\pi$.}
    \label{fig:g2_c2_exact}
\end{figure}

\clearpage

\subsection{Details on the theoretical model} \label{sec:theorysupmat}
For the results presented in Figure~\ref{fig:threshold_n}, we chose a Gaussian pulse shape for the coherent pump in Eq.~\eqref{eq:H}
 \begin{equation}
    \xi(t) = F_\text{pump} \frac{1}{\sqrt{2\pi \sigma^2}}\ee^{-\frac{(t-t_0)^2}{2\sigma^2}},\quad \sigma = \frac{\tau_\text{FWHM}}{2\sqrt{2\ln2}},\quad t_0 = \sigma\sqrt{-2\ln\epsilon}
\end{equation}
where the full-width-at-half-maximum $\tau_\text{FWHM}$ is defined by $\xi(t_0+\tau_\text{FWHM}/2) = \frac{\xi(t_0)}{2}$. The parameter $\epsilon$ defines the the pump strength at $t=0$ as $\xi(0) = \epsilon\xi(t_0)$, which we set to $\epsilon=0.001$, and $F_\text{pump}=\int_{-\infty}^\infty \xi(t) \dd t$ is the pump fluence.
Initially, the photons are in the vacuum state $\ket{0}$, all two-level systems are in the ground state $\ket{\g}$, and the vibrations are in a thermal state $\rho_{\text{vib},i}$
\begin{equation}
     \rho(0) = \dyad{0}\bigotimes_{i=1}^N(\dyad{\g}_i\otimes \rho_{\text{vib},i}),\quad \rho_{\text{vib},i} = (1-\ee^{-\omega_\nu/T})\ee^{-\omega_\nu b^\dagger_i b_i/T}.
\end{equation}
Before the pump pulse is applied, there is a delay to ensure that the molecular system consisting of the phonons and the electronic states has reached its steady state.

The second-order coherence at zero time delay $\tau$ and at time $t$ is defined by 
\begin{equation}
    g^{(2)}(t,\tau=0)\equiv g^{(2)}(t) = \frac{\langle a^\dagger(t) a^\dagger(t) a(t) a(t)\rangle}{\langle a^\dagger(t) a(t)\rangle ^2}.
\end{equation}
Since in the experiment an average value of $g^{(2)}(0)$ is extracted over the whole duration of the pulse, we define an averaged second-order coherence given by
\begin{equation}
    g^{(2)}_\text{avg}(0) = \frac{\int \langle a^\dagger(t) a^\dagger(t) a(t) a(t)\rangle \dd t }{\int \langle a^\dagger(t) a(t)\rangle ^2\dd t} \label{eq:g2_avg}
\end{equation}
Within mean-field theory, the expectation values factorize $\langle a^\dagger a^\dagger a a\rangle\approx \langle a^\dagger\rangle^2\langle a\rangle ^2$ and $\langle a^\dagger a\rangle\approx \langle a^\dagger\rangle\langle a\rangle$, such that $g^{(2)}_\text{MF}(t)=1$, which does not give us any information. Going one order higher, i.e., to a second-order cumulant approximation, one finds 
\begin{equation}
    \langle a^\dagger a^\dagger a a\rangle\approx 2\langle a^\dagger a\rangle^2  + \langle a^\dagger a^\dagger\rangle\langle a a\rangle - 2\langle a^\dagger\rangle^2\langle a\rangle ^2
\end{equation}
such that
\begin{equation}
    g^{(2)}_\text{c2}(t) = 2 + \frac{|\langle a(t) a(t)\rangle|^2 - 2|\langle a(t)\rangle|^4}{\langle a^\dagger(t) a(t)\rangle^2},
\end{equation}
where the subscript ``c2" indicates a second-order cumulant approximation. It is worth noting that $g^{(2)}_\text{c2}$ can only differ from two if symmetry-breaking cumulant equations are used, which do not assume that terms such as $\langle a\rangle$ vanish.
In this expression, one can again integrate the numerator and denominator to obtain an average second-order coherence.

For the results in Figure~\ref{fig:threshold_n}(c), we set the coherent pump to zero $\xi(t)=0$, and we impose initial conditions which resemble an excited ensemble of molecules by a coherent laser pulse \cite{freterTheoryDynamicalSuperradiance2025-si} 
\begin{equation}
     \rho(0) = \dyad{0}\bigotimes_{i=1}^N(\dyad{\psi(\theta)}_i\otimes \rho_{\text{vib},i}),\quad \ket{\psi(\theta)}_i =\cos(\theta/2)\ket{\text{e}}_i -\im\sin(\theta/2)\ket{\text{g}}_i.\label{eq:init}
\end{equation}
The angle $\theta$ determines the initial coherence in the system, and $\ket{\text g}_i, \ket{\text e}_i$ are the electronic ground state and excited state of the $i$th molecule. For $\theta=0$, the molecules are fully inverted, and there is no initial coherence in the system.

In Figure~\ref{fig:g2_c2_exact}, we benchmark the average second-order coherence calculated from second-order cumulants against a numerically exact quantum trajectories method~\cite{lloydPermutationSymmetricQuantum2026-si}. For this comparison, we excluded the vibrational degrees of  ($S=0$) and used an initial state as in Eq.~\eqref{eq:init} with $\theta=0.1\pi$. For large $N$, the cumulant result converges to the exact result.

\end{document}